\documentclass[10pt,twocolumn]{article}
\usepackage{amsmath,subfig,graphicx,cite,physics}
\usepackage[affil-it]{authblk}
\usepackage[margin=1.5cm]{geometry}
\usepackage{amsmath,amssymb}                                
\usepackage{makecell,multirow}
\usepackage{color}
\usepackage{textcomp}
\usepackage{algorithm,algpseudocode,algpascal}

\makeatletter
\renewcommand{\ALG@name}{Algorithm}
\makeatother

\begin{document}                                            

    \title{ Fast and robust method for flow analysis using GPU assisted diffractive optical element based background oriented schlieren (BOS)}
    \author[1]{Jagadesh Ramaiah}                               
    \author[2]{Sreeprasad Ajithaprasad}                        
    \author[1,2,5]{Gannavarpu Rajshekhar }          
    \author[3,4]{Dario Ambrosini}                             
    \affil[1]{Department of Electrical Engineering, Indian Institute of Technology Kanpur, Kanpur-208016, India }
    \affil[2]{Center for Lasers and Photonics, Indian Institute of Technology Kanpur, Kanpur-208016, India}
    \affil[3]{DIIIE, University of L’ Aquila, P. le E. Pontieri 1, L’ Aquila, 67100, Italy}
    \affil[4]{ISASI-CNR, Institute of Applied Science and Intelligent Systems, Via Campi Flegrei 34, Pozzuoli (NA), 80078, Italy}
\affil[5]{Email: gshekhar@iitk.ac.in}
\date{}
\maketitle
    \begin{abstract}                                            
        The paper introduces a method for studying flow dynamics using diffractive optical element based background-oriented schlieren (BOS).
        Our method relies on fringe demodulation using root multiple signal classification technique which provides high robustness against noise.
        Further, a graphics processing unit (GPU) based implementation is proposed which offers significant improvement in computational efficiency, and thus enables high speed analysis of flows.
        The performance of the method is demonstrated via numerical simulations and the practical applicability is also shown by analyzing a diffusion phenomenon in liquids by BOS. 
    \end{abstract}                                              

        {\bf Keywords:} Background-oriented schlieren, fringe projection, fringe analysis, phase retrieval, liquid diffusion

\section{Introduction}        
Study of the diffusion phenomena  \cite{cussler2009diffusion,tyrrell2013diffusion} is an important problem in flow analysis, and has applications in diverse areas such as biology \cite{yu2005diffusion}, mechanical and chemical engineering \cite{wijmans1995solution} and environmental sciences \cite{choy2017diffusion}.
An important class of techniques for diffusion measurements is comprised of the optical interferometric techniques, which offer several advantages such as full field measurement, good resolution, non-invasive operation and flexibility for both qualitative and quantitative assessment \cite{ambrosini2008overview,ambrosini2012}.  
Some of the prominent optical techniques in this domain include holographic interferometry \cite{gabelmann1979holographic,ruiz1985holographic,anand2006diffusivity,he2015development}, electronic speckle pattern interferometry \cite{paoletti1997temperature,riquelme2007interferometric,axelsson}, speckle decorrelation \cite{ambrosini2002speckle}, common path interferometry \cite{rashidnia2002development}, phase-shifting interferometry \cite{torres2012development} and projection moir\'e interferometry \cite{spagnolo2004liquid}. In particular, the method described by Spagnolo et al. \cite{spagnolo2004liquid} can be considered a type of background-oriented schlieren \cite{raffel2015background,settles2017review}.
The central idea in most of these techniques involves reliable extraction of phase distribution encoded in an interferogram or fringe pattern, since the phase distribution is directly mapped to the refractive index fluctuations caused by the diffusion process.
However, phase retrieval is not a trivial problem because of the challenges associated with noise, computational requirements associated with image size and total number of images, and single frame versus multi-frame operation and accordingly, several approaches have been proposed in literature.
Phase-shifting \cite{creath1985phase} is a popular multi-frame phase retrieval approach; however, the need for capturing multiple phase shifted interferograms poses technical difficulties for dynamics studies.
On the other hand, single-frame methods such as Fourier transform \cite{takeda1982fourier}, windowed Fourier transform \cite{kemao2007two,kemao2004windowed} and wavelet transform \cite{watkins1999determination,watkins2012review} can operate without the requirement of multiple fringe patterns.  
Among these methods, Fourier transform method has  been extensively applied in the domain of flow analysis and visualization \cite{spagnolo1994fourier,huntley01} mainly because of operational simplicity.  
However, Fourier transform is a global operation and its performance is severely affected by the presence of localized fringe abnormalities and noisy regions in the fringe pattern. 
For studying a dynamics based phenomenon such as diffusion, another important challenge is that a large number of time-lapsed fringe patterns are captured during imaging, and processing them for phase retrieval imposes significant computational burden.
This problem is even more apparent when the size of individual fringe pattern, measured in terms of number of pixels, is large.
Further, the investigation of fast diffusion process usually requires high speed imaging where the number of recorded fringe patterns is large to avoid any phase discontinuity between two consecutive images, and keep the phase variation small between successive fringe patterns.
In this case, the benefits of high performance fringe processing methods could be influential in the overall diffusion study.  

The main aim of our work is to propose a fringe processing method for flow analysis to address the twin challenges of noise sensitivity and computational efficiency.
Accordingly, in this paper, we propose root multiple signal classification  method for robust phase retrieval and show a highly efficient implementation using graphics processing unit based computing framework. 
In recent years, there is immense interest in the domain of fringe analysis for investigating methods with noise resistant capabilities \cite{montresor2016quantitative,montresor2019comparative,xia2018comparative,xia2017robust} and high performance operations \cite{gao2009real,van2016real,wang2019fast,vishnoi2019rapid}, and our work is a step in this direction. 
The outline of the paper is as follows.
The experimental technique based on diffractive optical element (DOE) background-oriented schlieren is described in section \ref{sec_exp}. 
The theory of the proposed method and the graphics processing unit based implementation is described in section \ref{sec_theory}.
The simulation and experimental results are described in section \ref{sec_results}.
This is followed by discussions and conclusions. 
Finally, in the appendix, some details are given about phase retrieval, refractive index variation and diffusion coefficient.

\section{Experimental setup}
\label{sec_exp}

\begin{figure}                                              
    \centering                                                  
    {\includegraphics[width=0.5\textwidth]{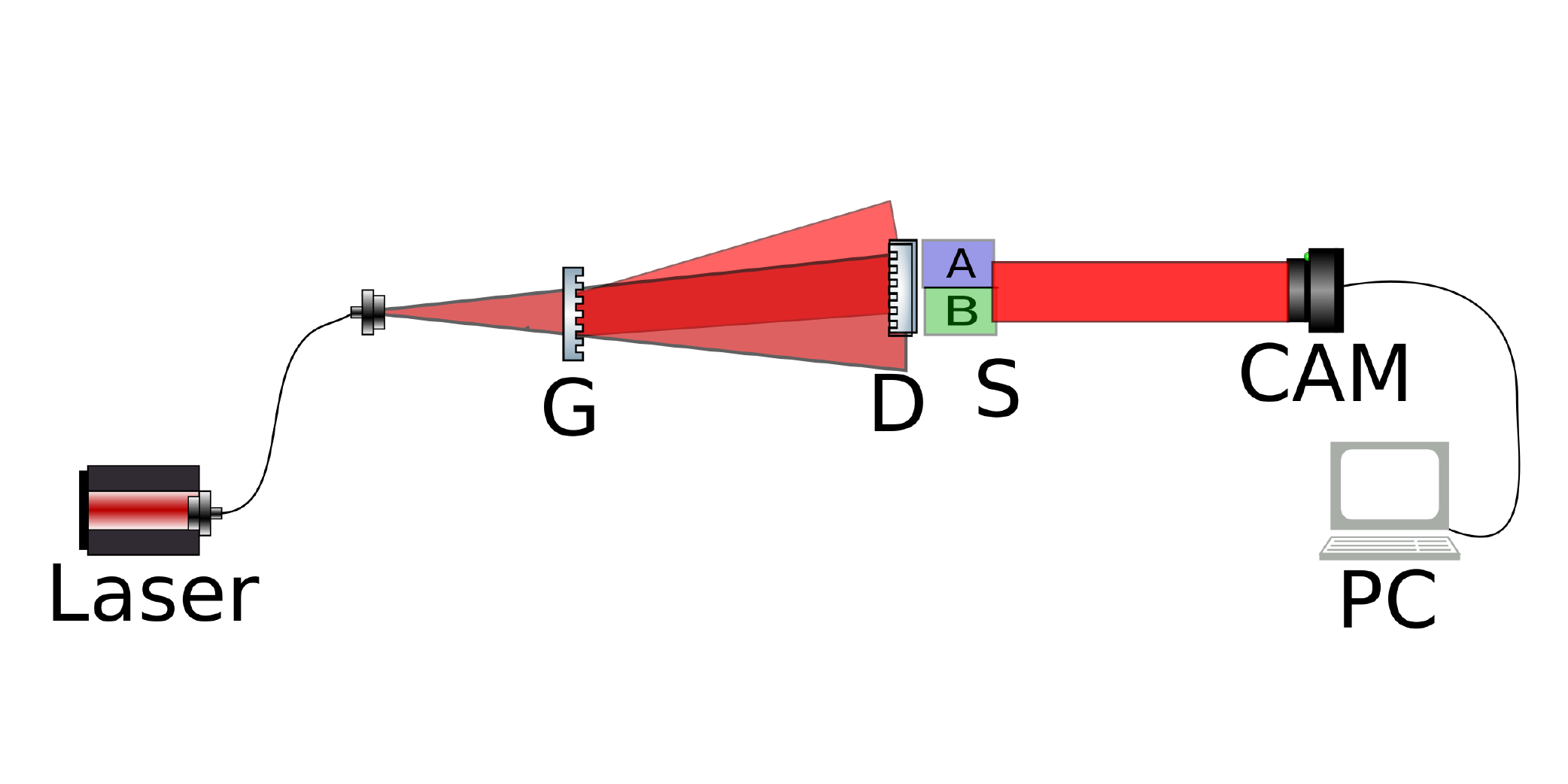}} \\
    \caption{Experimental setup.}                                
    \label{fig:exp_set}
\end{figure}

The schematic of the experimental setup used is shown in Fig. \ref{fig:exp_set}.
The experimental setup consist of a fiber coupled diode laser, diffractive optical element, grounded glass plate(D), diffusion cell and a camera. 
The diffractive element used in the system is a saw-tooth diffraction grating (G).
The diode laser produces coherent beam of wavelength $670$ nm and the spherical wave, originated from the tip of the single-mode optical fiber illuminates the saw-tooth grating. 
The diffraction efficiency for the $+1$ order and zeroth order of the grating is $0.31$ and $0.4$ respectively. 
Hence, most of the power incident on the grating(G) is divided between these two orders. 
Since, the intensity of these two orders are almost same, the interference pattern or grating pattern formed has high visibility.
This grating pattern is projected over a grounded glass plate (D) as shown in the schematic. 
The diffusive behavior of the grounded glass plate (D) makes the projected grating pattern visible at the camera plane. 
Binary liquid solution to be analyzed is kept inside a diffusion cell (S) made of glass and is placed after the grounded glass plate (D). 
The spectrophotometric glass cell, equipped with a Teflon shutting device to avoid evaporation phenomena, has internal dimensions of $10 \times 8$ mm$^2$, the path length along the optical axis is $10$ mm.
In the figure, A and B represent pure water and an aqueous solution of common salt, respectively. 

The diffusion cell is first half filled  with pure water (the solvent), then the solute (NaCl $1.75$ M/l aqueous solution) is injected from the bottom to reduce turbulence and mixing.
A CMOS TV camera (Silicon Video 9T001C with PIXCI® D2X imaging board by EPIX Inc., resolution 2048 $\times$ 1536 pixels, 3.2 $\mu$m $\times$ 3.2 $\mu$m pixel size), equipped with a TEC-55 $55$ mm F/$2.8$ Telecentric Computar Lens, captures the grating pattern at different times from the start of the diffusion process. The TEC-55 lens reduces viewing angle error and magnification error while providing good resolution and contrast with low distortion.
The refractive index distribution inside the diffusion cell is non-uniform, due to the concentration difference.  
Thus, the distorted grating patterns, as seen through the diffusion cell, encode the information about the refractive index variation and a set of sequentially captured patterns will contain the dynamics of the diffusion process. 
More details about the setup for analyzing the diffusion in liquids are outlined in \cite{spagnolo2004liquid}.

For our experiment, the recorded set of time-lapsed fringe patterns can be modeled as an image stack, where each image has a spatial carrier modulated cosinusoidal intensity variation.
Subsequently, by using bandpass filtering and carrier removal, the analytic or complex fringe signal \cite{spagnolo2004liquid,kemao2007two} is obtained as

\begin{equation}
    \Gamma(x,y,t) = A(x,y,t)e^{j\phi(x,y,t)} + \eta(x,y)
    \label{eqn:1}
\end{equation}
where $t$ indicates the time instant, $A$ indicates the amplitude and $\phi$ denotes the phase distribution induced by the refractive index fluctuations caused by diffusion process.  
Here, $\eta$ is the noise term which is assumed to be additive white Gaussian noise (AWGN).
As the cell dimensions remain unchanged throughout the experiment, the phase change with respect to time is primarily because of the refractive index variation induced by the diffusion process. 
Consequently, to investigate the dynamics of diffusion process, a reliable method for extracting the phase information is important. 
However, as mentioned before, phase estimation accuracy is severely deteriorated in the presence of noise.
Accordingly, to address this challenging problem, our proposed method is described in the next section.

\section{Theory}
\label{sec_theory}

In the proposed method, we select a region of the fringe signal by applying a moving
window of size $(2L + 1) \times (2L + 1)$ where $L$ is a length parameter. 
The moving window will slide over the fringe signal effectively creating multiple blocks.
Within the block, we assume the phase distribution to have a linear form and the fringe amplitude to have minimal variations so as to be approximately constant.
Subsequently, the fringe signal inside a block can be written as, 
\begin{equation}
    \boldsymbol{\Gamma}_w(x,y) = A_we^{j\phi_w(x,y)} + \eta_w(x,y)
    \label{eqn:wind}
\end{equation}
where the subscript "$w$" indicates the block index or number.
The linear phase $\phi_w$ inside the block can be modeled as
\begin{equation}
    \phi_w = \alpha + \omega_xx + \omega_yy
\end{equation}
As a consequence, the phase at any pixel can be calculated using Eq.(\ref{eqn:wind}) by estimating the unknowns $\alpha$, $\omega_x$ and $\omega_y$. We apply root multiple signal classification approach \cite{hayes2009statistical} to extract these parameters.

Consider the auto-correlation  of $\boldsymbol{\Gamma}_w$ denoted by $\boldsymbol{R}$ as,
\begin{equation}
    \boldsymbol{R} = E\{\boldsymbol{\Gamma}_w\boldsymbol{\Gamma}_w^H\} = \boldsymbol{R}_s + \boldsymbol{R}_n
    \label{eqn:autocor}
\end{equation}
where $E\{.\}$ and $(.)^H$ denote the expectation and conjugate transpose operations.
In Eq.(\ref{eqn:autocor}), $\boldsymbol{R}_s$ is called the signal auto-correlation matrix and is given as
\begin{equation}
    \boldsymbol{R}_s = A_w^2
    \begin{bmatrix}
        1 & e^{-j\omega_y} & \cdots & e^{-j(M-1)\omega_y} \\
        e^{j\omega_y} & 1 & \cdots & e^{-j(M-2)\omega_y} \\
        \vdots & \vdots & \ddots & \vdots \\
        e^{j(M-1)\omega_y} & e^{j(M-2)\omega_y} & \cdots & 1
    \end{bmatrix}
    \label{eqn:auto_matrix}
\end{equation}
and $\boldsymbol{R}_n = \sigma_w^2\boldsymbol{I}$ is the noise auto-correlation matrix with $\sigma_w^2$ denoting the noise covariance and $\boldsymbol{I}$ is an identity matrix of size $M\times M$ where $M=2L+1$.
Considering a vector defined as,
\begin{equation}
    \boldsymbol{u}_1 = 
    \begin{bmatrix}
        1 & e^{j\omega_y} & \cdots & e^{j(M-1)\omega_y}
    \end{bmatrix}^T
\end{equation}
Eq.(\ref{eqn:auto_matrix}) can be simplified as
\begin{equation}
    \boldsymbol{R}_s = A_w^2\boldsymbol{u}_1\boldsymbol{u}_1^H
\end{equation}
Post multiplying the vector $\boldsymbol{u}_1$ on both sides of the above equation, we get
\begin{equation}
    \begin{aligned}
        \boldsymbol{R}_s\boldsymbol{u}_1 &= A_w^2\boldsymbol{u}_1(\boldsymbol{u}_1^H\boldsymbol{u}_1) \\
        &= (MA_w^2)\boldsymbol{u}_1
        \label{eqn:eigen}
    \end{aligned}
\end{equation}
which implies that the vector $\boldsymbol{u}_1$ is an eigenvector of $\boldsymbol{R}_s$ corresponding to the eigenvalue equal to $MA_w^2$.
Since $\boldsymbol{R}_s$ is a Hermitian matrix, the remaining eigenvectors, $\boldsymbol{u}_2$, $\boldsymbol{u}_3$, $\cdots$, $\boldsymbol{u}_M$ will be orthogonal to $\boldsymbol{u}_1$.
\begin{equation}
    \boldsymbol{u}_1^H\boldsymbol{u}_i = 0\quad;\quad i = 2,3,\cdots,M
    \label{eqn:ortho}
\end{equation}
If $\lambda_i^s$ are the eigenvalues of $\boldsymbol{R}_s$, then Eq. \ref{eqn:autocor} becomes
\begin{equation}
    \begin{aligned}
        \boldsymbol{R}\boldsymbol{u}_i &= (\boldsymbol{R}_s + \sigma_w^2\boldsymbol{I})\boldsymbol{u}_i \\
        &= (\lambda_i^s+ \sigma_w^2)\boldsymbol{u}_i
    \end{aligned}
    \label{eqn:eigen_val}
\end{equation}
Thus the eigenvalues of noisy data matrix $\boldsymbol{\Gamma}_w$ are $\lambda_i = \lambda_i^s+ \sigma_w^2$.
From equations (\ref{eqn:eigen}) and (\ref{eqn:eigen_val}), it is clear that $\boldsymbol{u}_1$ is an eigenvector corresponding to the largest eigenvalue given by $\lambda_1 = MA_w^2+\sigma^2$.
The vector $\boldsymbol{u}_1$ represents the signal subspace \cite{stoica2005spectral}.
Similarly, the matrix containing all the remaining eigenvectors represents noise subspace and is given as,
\begin{equation}
    \boldsymbol{U}_n = 
    \begin{bmatrix}
        \boldsymbol{u}_2 & \boldsymbol{u}_3 & \cdots & \boldsymbol{u}_M
    \end{bmatrix}_{M\times (M-1)}
\end{equation}
From Eq.(\ref{eqn:ortho}), we can infer that the signal subspace and noise subspace are orthogonal to each other.
Using this property, the unknown $\omega_y$ can be estimated by solving the polynomial equation  given as \cite{stoica2005spectral}
\begin{equation}
    \lvert \boldsymbol{u}_1^H\boldsymbol{U}_n \rvert^2 = \boldsymbol{u}_1^H(z_y)\boldsymbol{U}_n\boldsymbol{U}_n^H\boldsymbol{u}_1(z_y) = 0
    \label{eqn:poly1}
\end{equation}
where 
$
\boldsymbol{u}_1(z_y) = 
\begin{bmatrix}
    1 & z_y & \cdots & z_y^{M-1}
\end{bmatrix}^T
$ and $z_y = e^{j\omega_y}$.
Following the similar analysis for the auto-correlation matrix given by $E\{\boldsymbol{\Gamma}_w^H\boldsymbol{\Gamma}_w\}$, we get a polynomial equation in $\omega_x$ as,
\begin{equation}
    \lvert \boldsymbol{v}_1^H\boldsymbol{V}_n \rvert^2 = \boldsymbol{v}_1^H(z_x)\boldsymbol{V}_n\boldsymbol{V}_n^H\boldsymbol{v}_1(z_x) = 0
    \label{eqn:poly2}
\end{equation}
where 
$
\boldsymbol{v}_1(z_x) = 
\begin{bmatrix}
    1 & z_x & \cdots & z_x^{M-1}
\end{bmatrix}^T
$ represent signal subspace with $z_x = e^{-j\omega_x}$ and the noise subspace is given by,
\begin{equation}
    \boldsymbol{V}_n = 
    \begin{bmatrix}
        \boldsymbol{v}_2 & \boldsymbol{v}_3 & \cdots & \boldsymbol{v}_M
    \end{bmatrix}_{M\times (M-1)}
\end{equation}

For our analysis, the eigenvectors of the auto-correlation matrices were computed using the singular value decomposition approach \cite{golub2012matrix}.
Further, the polynomial equations in Eq. \ref{eqn:poly1} and \ref{eqn:poly2} can be solved using eigenvalue decomposition of the companion matrix as discussed in \cite{chapra2012applied}.
Among the $(2M-1)$ possible roots, we select the one which is closest to the unit circle with magnitude less than 1 on the complex plane as the roots of $z_y$ and $z_x$.
Finally, the unknown parameters are estimated by solving Eqns.(12) and (13) for $z_y$ and $z_x$, and are given as,
\begin{equation}
    \begin{aligned}
        \omega_y &= \textit{arg}(z_y) \\
        \omega_x &= -\textit{arg}(z_x) \\
        \alpha &= \angle \left[\overline{\boldsymbol{\Gamma}_we^{-j(\omega_xx+\omega_yy)}}\right]
    \end{aligned}
\end{equation}
where $\angle(\cdot)$ denotes the angle operation and $\overline{(.)}$ denotes mean operation.
The above process is repeated for all blocks and followed by an unwrapping operation \cite{herraez2002fast} to obtain the overall phase map.
The main advantage of the proposed approach is that the signal and noise components can be effectively separated as discussed above which provides high robustness against noise.
However, the matrix based operations associated with the proposed approach are usually computationally intensive.

Hence, to improve the computational efficiency, the proposed method is implemented using graphics processing unit (GPU) \cite{nvidia2011nvidia, sanders2010cuda}.  
Graphics processing unit is a specialized hardware designed for parallel execution of a given task on numerous threads and is rapidly emerging as a powerful tool for parallel computing in optical metrology \cite{gao2012parallel,wang2018parallel}.
Any recursive and independent tasks can be effectively performed on GPU, resulting in high computation speed.
In general, a GPU is connected to a central processing unit (CPU) called the host computer which controls the execution of GPU threads.
A C/C++ based heterogeneous programming model called compute unified device architecture (CUDA) was created by NVIDIA to program both host computers and CUDA enabled GPUs.
The threads are grouped into multiple blocks so that each block is executed concurrently based on the availability of streaming multiprocessors.
The code for host CPU is written in C language like functions which are executed sequentially.
But the task required to be executed in parallel on GPU is written using a special function called kernel.
The user can also specify the number of threads and blocks to be launched while calling a kernel function.
To implement the proposed method, we used NVIDIA's Quadro-M5000 GPU which has 16 streaming multiprocessors and supports up to one thousand threads in each block.
Also, the native single precision CUDA arithmetic was used in our computations.
As the proposed method is based on independent pixel wise operations, the processing of each window can be programmed using a GPU kernel function so that each thread can estimate the phase at the corresponding pixel.
This GPU computing based approach has the potential to significantly speed up the computations required in the proposed method.

\begin{algorithm}[t]
\caption{Pseudo-code of GPU kernel used for phase estimation}\label{Alg:kernel}
\begin{algorithmic}[1]
\Function gpu\_kernel (int $N_y$, int $N_x$)
\State \quad (px, py) $\leftarrow$ Compute pixel indexes
\State \quad \textbf{if} (px $<$ $N_x$ and py $<$ $N_y$)
\State \qquad $\boldsymbol{\Gamma}_w$ $\leftarrow$ $(M\times M)$ window around (px, py)
\State \qquad \textbf{U},\textbf{S},$\textbf{V}^H$ $\leftarrow$ SVD of $\boldsymbol{\Gamma}_w$
\State \qquad $\textbf{U}_n$ $\leftarrow$ $\begin{bmatrix}
\boldsymbol{u}_2 & \boldsymbol{u}_3 & \cdots & \boldsymbol{u}_M
\end{bmatrix}_{M\times (M-1)}$
\State \qquad $\textbf{V}_n$ $\leftarrow$ $\begin{bmatrix}
\boldsymbol{v}_2 & \boldsymbol{v}_3 & \cdots & \boldsymbol{v}_M
\end{bmatrix}_{M\times (M-1)}$
\State \qquad y\_poly $\leftarrow$ $\boldsymbol{u}_1^H(z_y)\boldsymbol{U}_n\boldsymbol{U}_n^H\boldsymbol{u}_1(z_y)$
\State \qquad x\_poly $\leftarrow$ $\boldsymbol{v}_1^H(z_x)\boldsymbol{V}_n\boldsymbol{V}_n^H\boldsymbol{v}_1(z_x)$
\State \qquad $z_y$ $\leftarrow$ Root of y\_poly which is inside and closest to unit circle.
\State \qquad $z_x$ $\leftarrow$ Root of x\_poly which is inside and closest to unit circle. 
\State \qquad $\phi(px,\ py)$ $\leftarrow$ Compute phase using Eq. (15)
\State \quad  \textbf{end if}
\State \textbf{end function}
\end{algorithmic}
\end{algorithm}

Algorithm 1 lists the step-wise operations performed by the gpu\_kernel function for estimating phase using the proposed method.
When this kernel is called by CPU, the GPU launches an user specified number of threads in parallel where each thread simultaneously performs the steps listed in Pseudo-code 1.
The pixel index of a particular thread can be calculated using CUDA specific variables that are accessible in each active thread.
Note that in steps 6 and 7, the vectors $\boldsymbol{u}_i$ and $\boldsymbol{v}_i$ represent $i^{th}$ column of $\boldsymbol{U}$ and $\boldsymbol{V}$ respectively.
After computing the polynomial roots, the estimated phase at the pixel corresponding to the thread is calculated using Eq. (15).

\section{Results}
\label{sec_results}
To demonstrate the applicability of the proposed method, we simulated a noisy complex fringe signal at signal to noise ratio (SNR) equal to 0 dB.
The size of the simulated fringe pattern is 512 $\times$ 512 pixels. 
All simulations were performed using Numpy \cite{van2011numpy}, which is an efficient scientific library for the Python programming language.
The main advantage of this library is that it provides an easy implementation for multi-dimensional array or matrix datatype similar to MATLAB, but is purely open source and free of cost.  

\begin{figure}[t!]
    \centering
    {\includegraphics[width=0.5\textwidth]{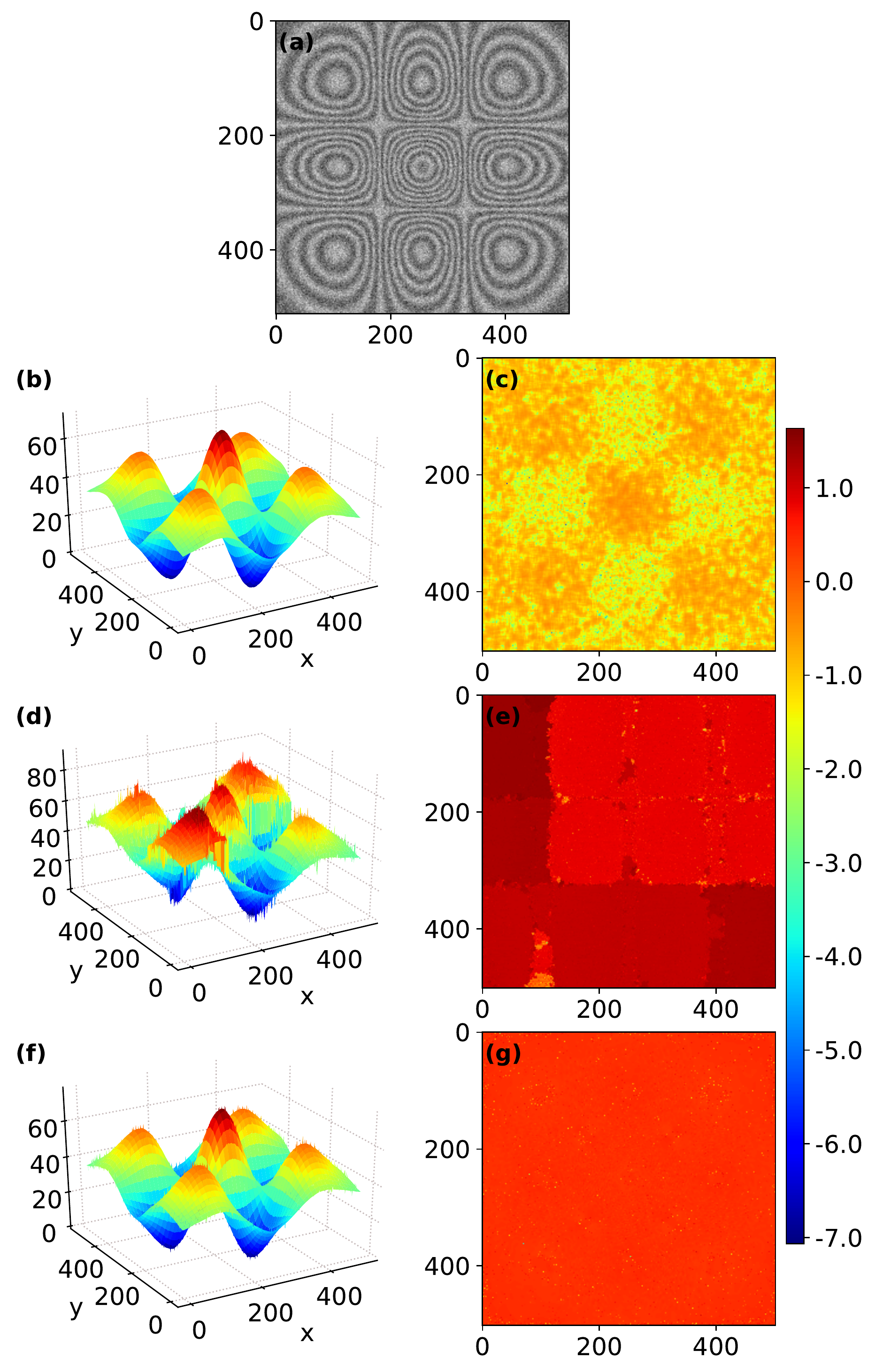}}
    \caption{(a) Simulated fringe pattern of size $512\times 512$ at SNR of 0 dB.
    Estimated phases in radians using (b) proposed method, (d) 1D-WT method and (f) 2D-WT method.
    Corresponding absolute phase estimation errors (logarithmic values) are shown in (c), (e) and (g).}
    \label{fig:sim_results}
\end{figure}

The real part of the signal or the fringe pattern is shown in Fig.\ref{fig:sim_results}(a).
Subsequently, we applied the proposed method for phase retrieval from the noisy fringe signal.
The estimated phase in radians using the proposed method with $L=5$ is shown in Fig.\ref{fig:sim_results}(b) and the corresponding absolute phase estimation error ( logarithmic values) is shown in Fig.\ref{fig:sim_results}(c)
For comparison, the estimated phases using both one dimensional wavelet transform (1D-WT)\cite{watkins1999determination} and two dimensional wavelet transform (2D-WT) \cite{watkins2012review} are shown in figures \ref{fig:sim_results}(d) and (f).
The wavelet transform based approach is popular for analyzing a non-stationary signal such as fringe pattern. 
It offers an excellent tool for studying the local spatial frequencies associated with the fringe pattern and provides multi-resolution space-frequency analysis capability \cite{watkins2012review}. 
The corresponding estimation errors (logarithmic values) using the wavelet transform methods are also given in Fig.\ref{fig:sim_results}(e) and (g). 
Note that border pixels are neglected to ignore the errors at the boundaries.

\begin{figure}[t!]
    \centering
    {\includegraphics[width=0.5\textwidth]{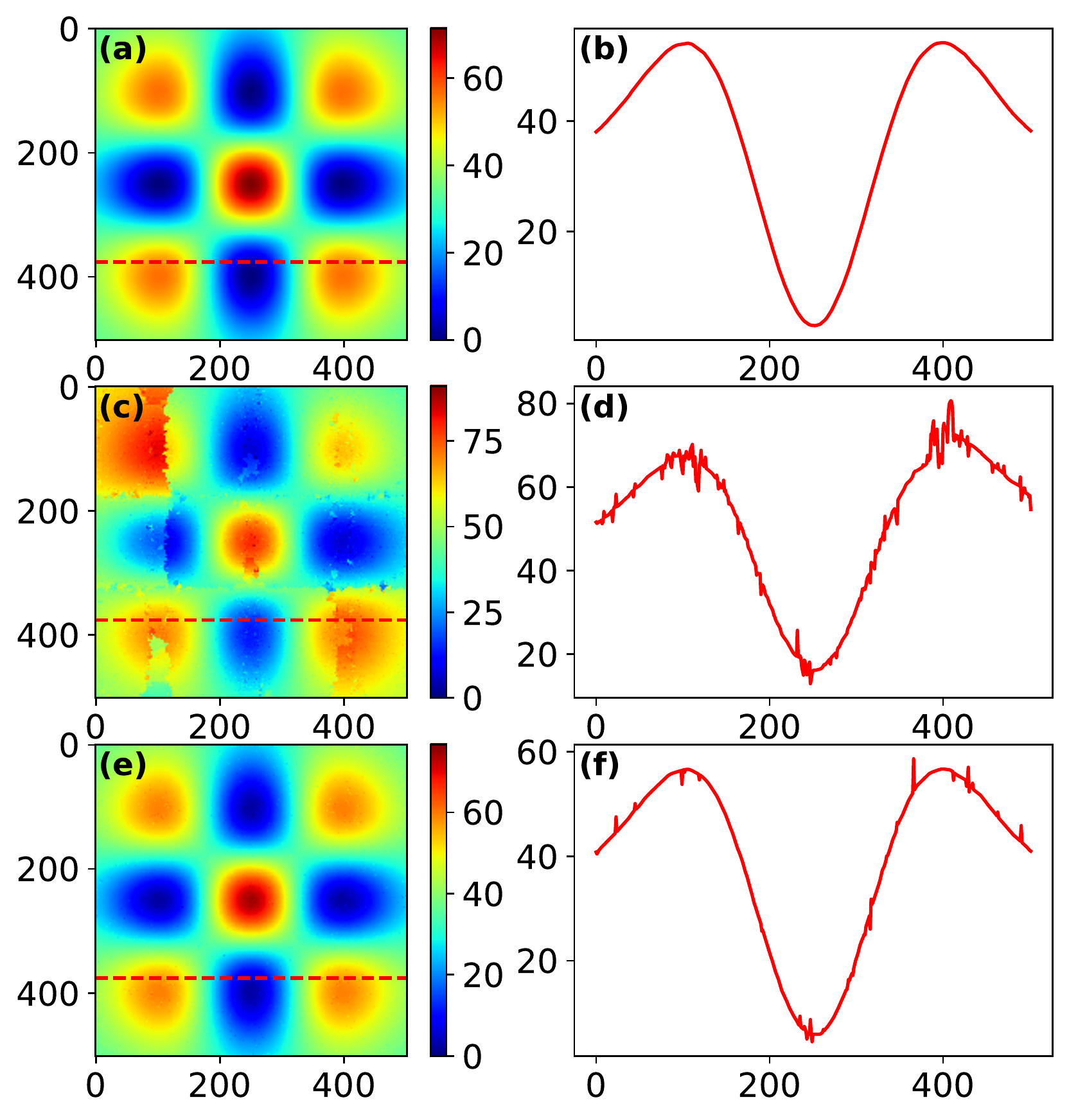}}
    \caption{Top view of estimated phase using (a) proposed method, (c) 1D-WT method and (e) 2D-WT method with the corresponding line profiles along the dashed lines shown in (b), (d) and (f) respectively.}
    \label{fig:sim_results_3}
\end{figure}

\begin{figure}[t!]
    \centering
    {\includegraphics[width=0.5\textwidth]{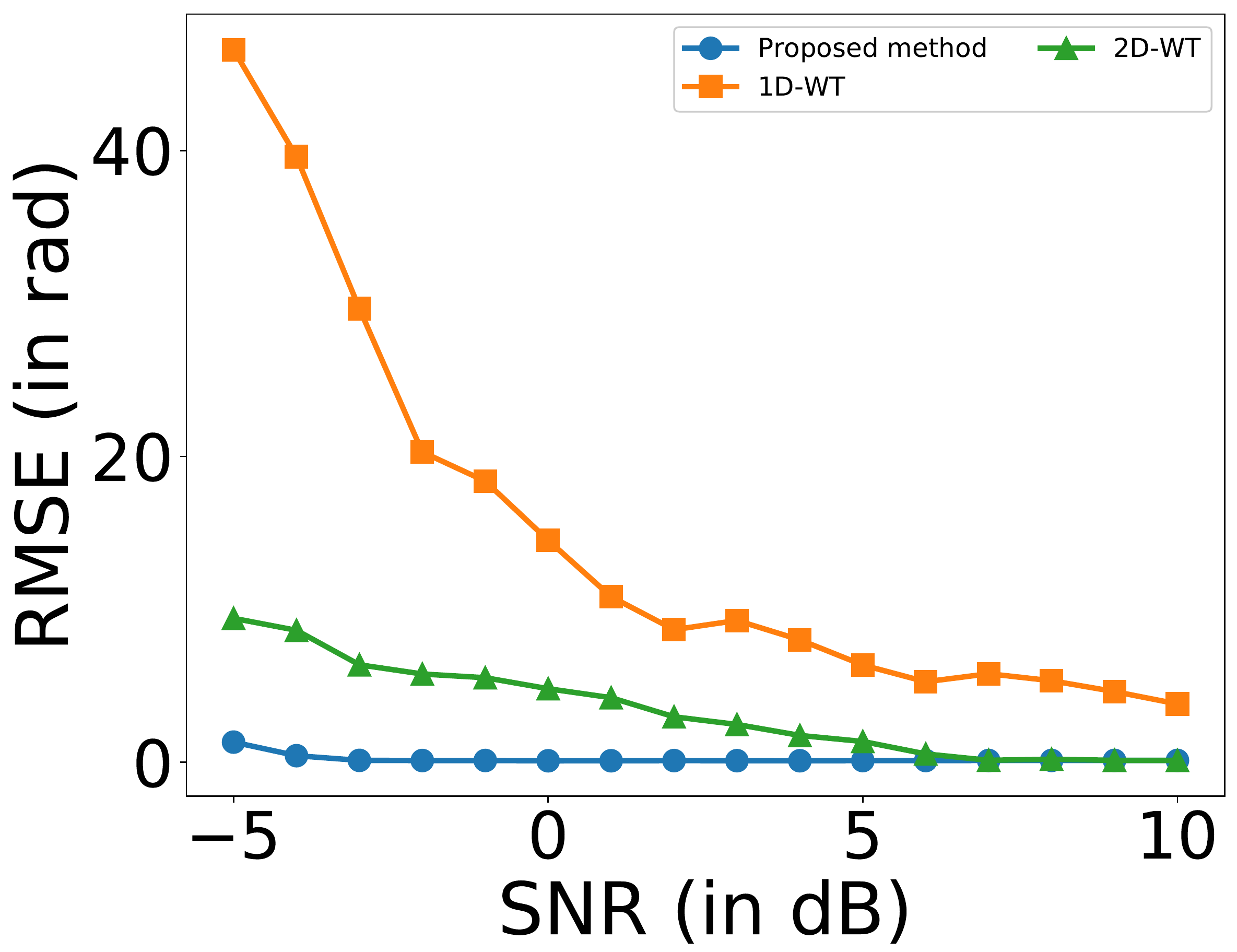}} \\
    \caption{Plot comparing the RMSE of the estimated phase using various methods.}
    \label{fig:sim_rmse}
\end{figure}

For better visualization, we also show the top view images of the estimated phases in radians using the proposed method, one-dimensional wavelet transform method and two-dimensional wavelet transform method in parts (a,c,e) of Figure \ref{fig:sim_results_3}.
Without loss of generality, an arbitrary row (marked by red dashed line) was chosen in the top view images, and the corresponding line profile plots of the estimated phases in radians are shown in parts (b,d,f) of Figure \ref{fig:sim_results_3}.
It is evident that the proposed method offers a smoother line profile as compared to the other methods.

For quantitative assessment, we also computed the root mean square errors (RMSE) for phase estimation using the different methods at various levels of noise.
The RMSE versus SNR plot for the proposed method and the wavelet based methods is shown in Fig.\ref{fig:sim_rmse}.
This figure clearly highlights the robustness of the proposed method against noise for fringe processing.

\begin{table}[t]
\centering
\caption{Comparison of phase estimation errors for different window sizes}
\label{tab:1}
\begin{tabular}{|c|c|}
\hline
\makecell{L (in pixels)} & \makecell{RMSE (in radians)}\\ \hline
  1 & 5.3823\\ \hline
  2 & 0.4893\\ \hline
  3 & 0.1090\\ \hline
  4 & 0.0945\\ \hline
  5 & 0.1316\\ \hline
  6 & 0.1552\\ \hline
  7 & 0.2185\\ \hline
  8 & 0.2937\\ \hline
\end{tabular}
\end{table}

Further, we also investigated the effect of window size on estimation accuracy.
In Table \ref{tab:1}, the root mean square errors for phase estimation are displayed for varying values of the window size related parameter $L$.
For a small window, the linear phase approximation is more valid; however, this brings high noise susceptibility since less number of data samples are processed.
On the other hand, a larger window size offers better robustness against noise, though the accuracy of linear phase model deteriorates in this case.

With respect to computational efficiency, as mentioned before, we implemented the proposed method using graphics processing unit.
For comparison, we also implemented  C programming language based sequential processing using gcc compiler with optimization option (-O3) enabled.   
The comparison between the sequential processing versus the parallel GPU computing approach is shown in Table \ref{tab:2}.
It is evident that as the image size grows, tremendous improvements in execution runtime can be achieved using the graphics processing unit based implementation.

\begin{table}[t]
    \centering
    \caption{Comparison of computation time for estimating phase from fringe patterns of varied sizes.}
    \label{tab:2}
    \begin{tabular}{|c|c|c|}
        \hline
        \makecell{} & \multicolumn{2}{c|}{\makecell{Execution time (sec)}} \\ \cline{2-3}
        \makecell{Image size\\ (in pixels)} & \makecell{CPU\\(using C)} & \makecell{GPU\\(using CUDA)}  \\ \hline
        $256\times 256$ &   30.09  &  1.02  \\ \hline
        $512\times 512$ &  121.30  &  3.59 \\ \hline
        $1024\times 1024$ &  486.81  & 13.88 \\ \hline
        $2048\times 2048$ & 1952.65  & 55.36 \\ \hline
    \end{tabular}
\end{table}

\begin{figure}[t!]
    \centering
    {\includegraphics[width=0.5\textwidth]{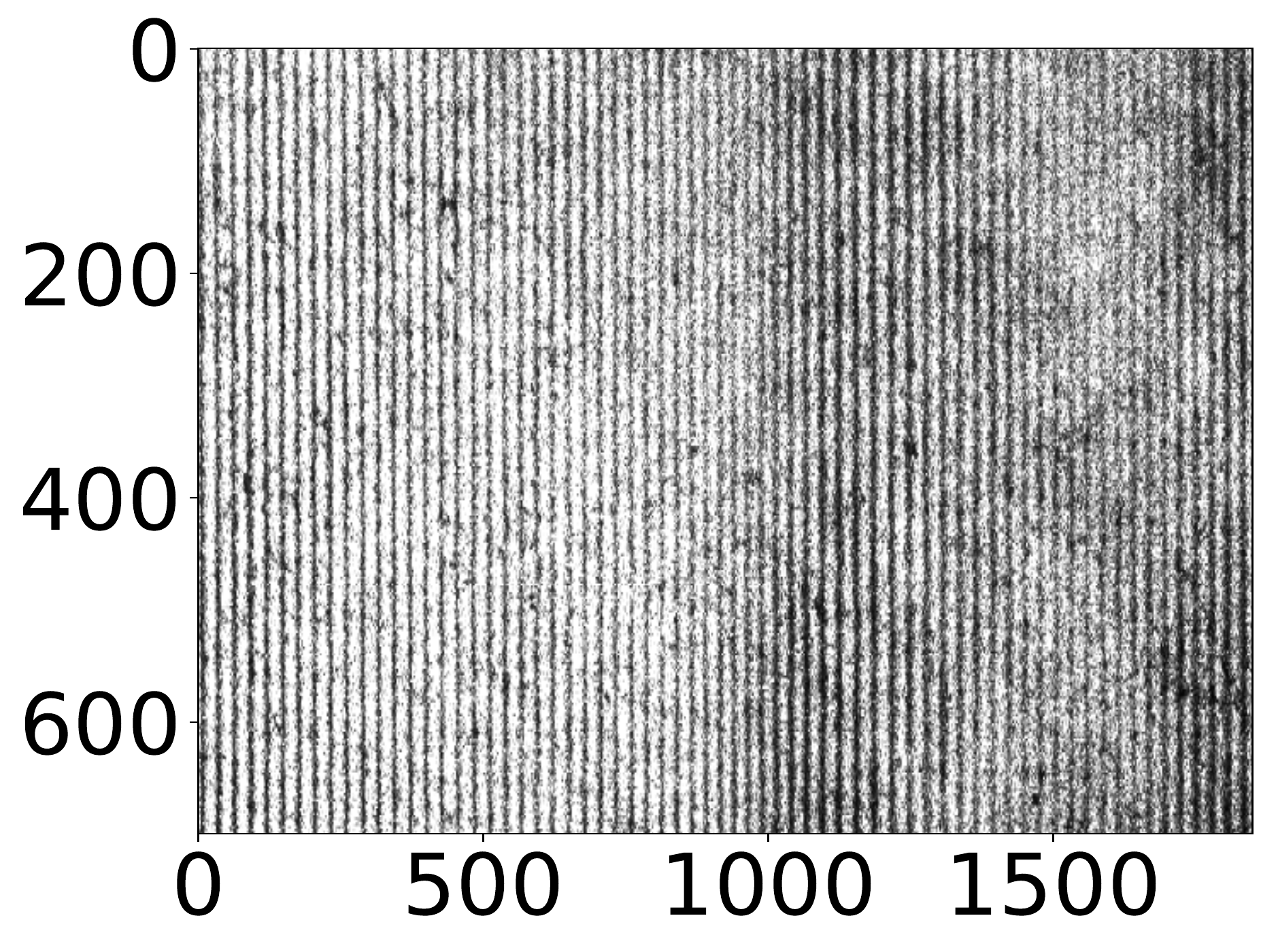}}
    \caption{An experimentally recorded fringe pattern.}
    \label{fig:exp_results_4}
\end{figure}

\begin{figure}[t!]
    \centering
    {\includegraphics[width=0.5\textwidth]{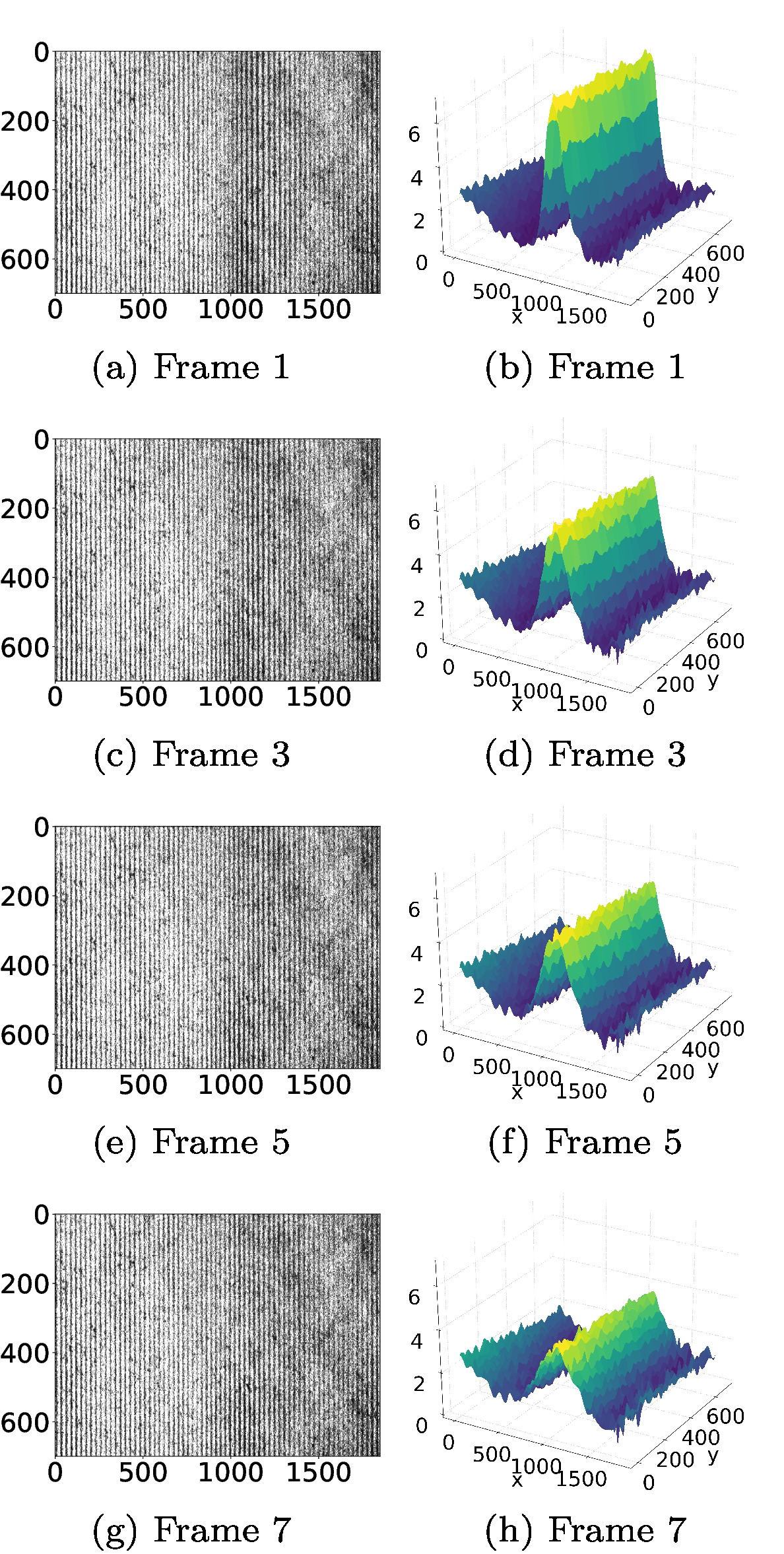}}
    \caption{(a,c,e,g) Experimentally recorded fringe patterns and (b,d,f,h) corresponding phase estimated using the proposed method.}
    \label{fig:exp_results_1}
\end{figure}

\begin{figure}[t!]
    \centering
    {\includegraphics[width=0.5\textwidth]{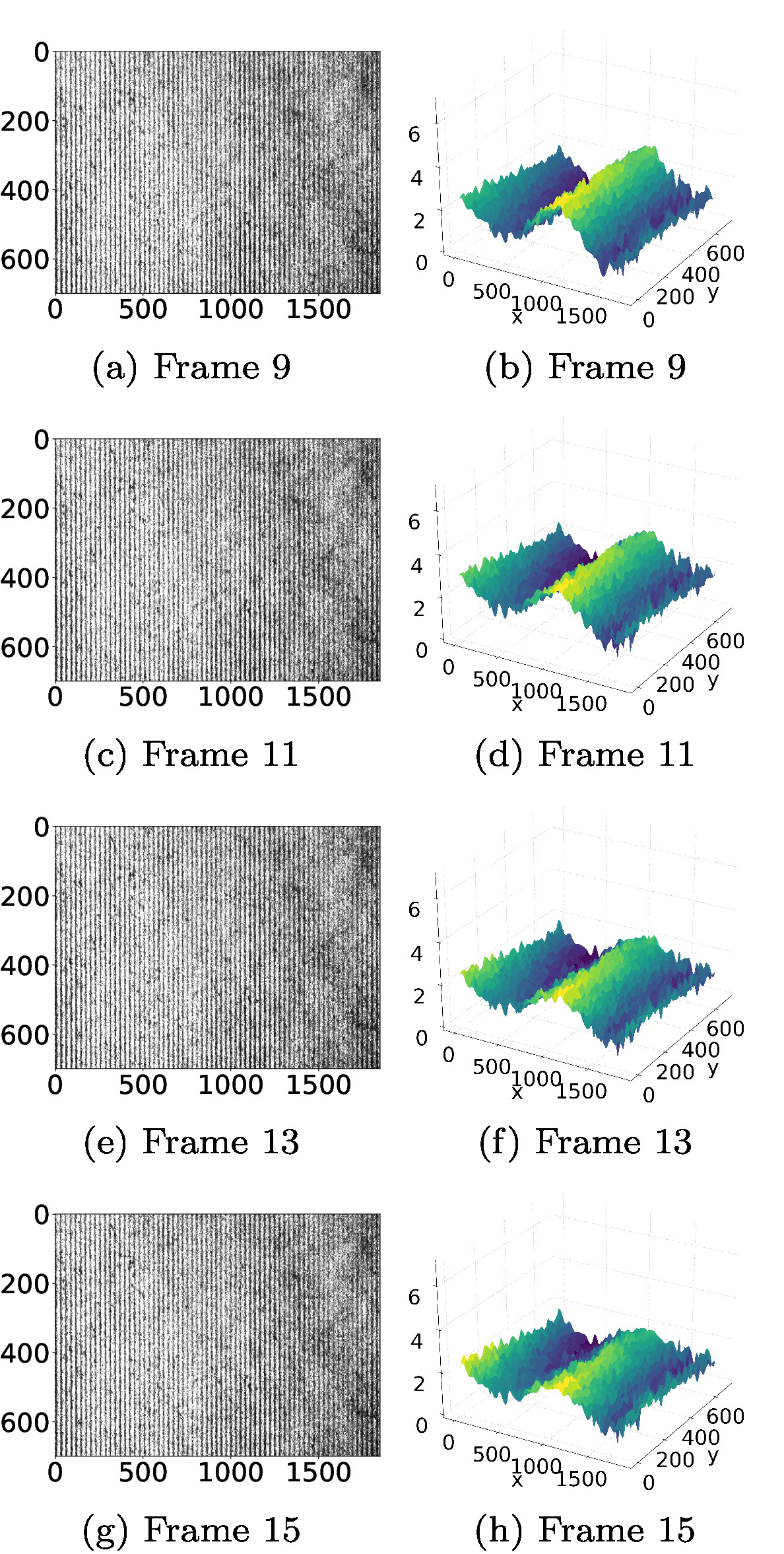}}
    \caption{(a,c,e,g) Experimentally recorded fringe patterns and (b,d,f,h) corresponding phase estimated using the proposed method.}
    \label{fig:exp_results_2}
\end{figure}

\begin{figure}[t!]
    \centering
    {\includegraphics[width=0.5\textwidth]{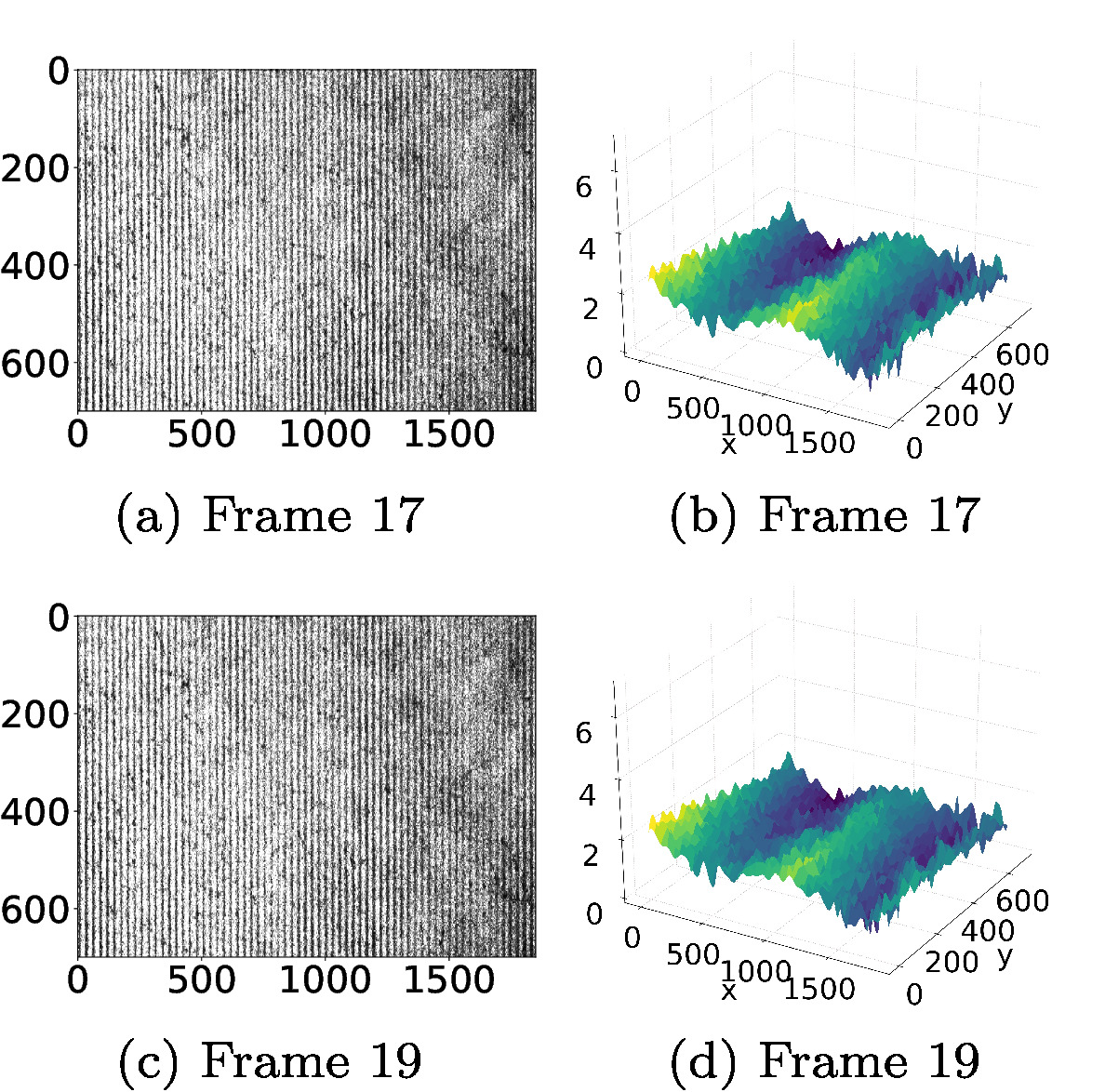}}
    \caption{(a,c) Experimentally recorded fringe patterns and (b,d) corresponding phase estimated using the proposed method.}
    \label{fig:exp_results_3}
\end{figure}

The utility of the proposed method for practical applications is demonstrated using the experimentally recorded fringe patterns in diffractive optical element based background-oriented schlieren. Images (8-bit depth) were taken using a f-number = 4,  an exposure time of $2.5$ ms and with a magnification such as $1$ pixel = $9.1$ $\mu$m.
The time-lapsed data set comprises of 19 sequentially captured fringe patterns for the diffusion experiment. 
Frames 1-5 were recorded at time intervals of $120$ s (i.e. $120$ s, $240$ s, $360$ s, $480$ s and $600$ s from the beginning of diffusion) while Frames 6-19 were recorded at time intervals of $300$ s (i.e. $900$ s, $1200$ s, $1500$ s etc.).
For experimental analysis, fringe patterns with size of 1850 $\times$ 700 pixels were processed using the proposed method for phase recovery with $L=7$.
A representative fringe pattern from the data set is shown in Figure \ref{fig:exp_results_4}.

The images marked by different frame numbers are shown in parts (a,c,e,g) of Figures \ref{fig:exp_results_1}, \ref{fig:exp_results_2} and  parts (a,c) of Figure \ref{fig:exp_results_3}.
After phase unwrapping, the estimated phases in radians using the proposed method for different fringe patterns are shown in parts (b,d,f,h) of Figures \ref{fig:exp_results_1}, \ref{fig:exp_results_2}, and parts (b,d) of Figure \ref{fig:exp_results_3}.

In these figures, the variation in phase distribution with time is clearly evident.
This is because as the diffusion process progresses with time, the initially non-uniform refractive index distribution of the mixture proceeds towards a uniform distribution.
These temporal fluctuations in refractive index lead to equivalent temporal phase variations exhibited in the shown images.

\section{Discussion}
Diffractive optical element based BOS provides a simple and unsophisticated technique for studying flow based phenomenon such as diffusion and can be easily applied by unskilled operator.
The applicability of this system for flow analysis can be tremendously improved by the infusion of robust data processing methods.
Our work is a step in this direction by proposing a noise resistant and fast method for fringe processing.
The performance of the proposed method is validated using the shown simulation and experimental results.
Further, the applied GPU computing methodology provides significant improvements in computational efficiency, and paves path for rapid flow analysis using optical interferometry.
The utility of this framework is especially evident when large number of fringe patterns and with big image sizes need to be processed. 
In addition, the external factors which deteriorate the estimation accuracy in fringe processing include temperature effects, vibrations and air currents. 
To an extent, their effect is addressed by the single shot fringe processing operation of the proposed method where only one frame is required for extracting the desired phase map.
In contrast,  multi-frame methods such as phase-shifting algorithms require the recording of multiple fringe patterns, which can induce more susceptibility to these external disturbances. For more details about accuracy and reproducibility in diffusion measurements, the Reader is addressed to \cite{axelsson} and references therein.

\section{Conclusions}
In this paper, we proposed an elegant fringe processing method in DOE based background-oriented schlieren for flow analysis.
The dual advantages of high noise robustness and good computational efficiency attributed to graphic processing unit based computing provide great practical utility to the proposed method.
The authors believe that the proposed method has great potential for flow analysis and visualization.

\section*{Acknowledgments}
Dr. Gannavarpu Rajshekhar gratefully acknowledges the funding obtained from Department of Science and Technology (DST), India under grant number DST/NM/NT/2018/2.

\section*{Appendix}
This section is about phase retrieval, refractive index retrieval and diffusion coefficient. These topics are discussed in depth in literature, e.g. \cite{ambrosini2008overview,gabelmann1979holographic,ruiz1985holographic,riquelme2007interferometric,axelsson, rashidnia2002development,spagnolo2004liquid}, therefore only a very brief account will be given here for the coherence of the paper.

"Diffusion is the process by which matter is transported from one part of a system to another as a result of random molecular motion" \cite{crank}; the final result of the process is complete mixing. A free diffusion process can be described by Fick's second law: for a one dimensional diffusion
\begin{equation}
    \pdv{c(x,t)}{ t} =D \pdv[2]{c(x,t)}{x}
    \label{eqn:diff1}
\end{equation}
where $c$ is the concentration, $x$ and $t$ are position and time, respectively  and $D$ is the diffusion coefficient (independent from $c$). For dilute solutions, the refractive index $n$ can be treated as a linear function of the concentration $c$. The solution of Eq.(\ref{eqn:diff1}) for two binary liquid mixtures, initially ($t=0$) separated at $x=0$, is available in literature \cite{ambrosini2008overview,gabelmann1979holographic,spagnolo2004liquid,crank}. A change in concentration induces a change in refractive index; this non uniform $n$ deflects the beam travelling through the test section \cite{spagnolo2004liquid}. The change of the beam deflection angle can be considered as a local shift of the fringe pattern as seen through the diffusion cell, giving rise to the phase term $\phi$ of Eq.(\ref{eqn:1}). In particular, under paraxial approximation, the expression relating the refractive index derivative to the phase term can be written as \cite{spagnolo2004liquid}
\begin{equation}
    \pdv{n(x,t)}{x} =\frac{1}{2\mu f_x}\frac{n_0}{L^2} \phi(x,t)
    \label{eqn:diff2}
\end{equation}
where $f_x$ is the fringe frequency, $n_0$ is a suitable index of refraction and $L$ represents the test cell dimension along the propagation axis. Therefore the estimated phase, presented in parts (b,d,f,h) of Figures \ref{fig:exp_results_1}, \ref{fig:exp_results_2}, and parts (b,d) of Figure \ref{fig:exp_results_3} are proportional to the derivative of $n$.


\begin{thebibliography}{10}
\expandafter\ifx\csname url\endcsname\relax
  \def\url#1{\texttt{#1}}\fi
\expandafter\ifx\csname urlprefix\endcsname\relax\def\urlprefix{URL }\fi
\expandafter\ifx\csname href\endcsname\relax
  \def\href#1#2{#2} \def\path#1{#1}\fi

\bibitem{cussler2009diffusion}
E.~L. Cussler, Diffusion: mass transfer in fluid systems, Cambridge University
  Press, 2009.

\bibitem{tyrrell2013diffusion}
H.~J.~V. Tyrrell, K.~Harris, Diffusion in liquids: a theoretical and
  experimental study, Butterworth-Heinemann, 2013.

\bibitem{yu2005diffusion}
H.~Yu, I.~Meyvantsson, I.~A. Shkel, D.~J. Beebe, Diffusion dependent cell
  behavior in microenvironments, Lab on a Chip 5~(10) (2005) 1089--1095.

\bibitem{wijmans1995solution}
J.~G. Wijmans, R.~W. Baker, The solution-diffusion model: a review, Journal of
  membrane science 107~(1-2) (1995) 1--21.

\bibitem{choy2017diffusion}
B.~Choy, D.~D. Reible, Diffusion models of environmental transport, CRC Press,
  2017.

\bibitem{ambrosini2008overview}
D.~Ambrosini, D.~Paoletti, N.~Rashidnia, Overview of diffusion measurements by
  optical techniques, Optics and Lasers in Engineering 46~(12) (2008) 852--864.

\bibitem{ambrosini2012}
D.~Ambrosini, J.-P. Prenel, Flow visualization and beyond,
  Optics and Lasers in Engineering 50 (2012) 1-7.

\bibitem{gabelmann1979holographic}
L.~Gabelmann-Gray, H.~Fenichel, Holographic interferometric study of liquid
  diffusion, Applied Optics 18~(3) (1979) 343--345.

\bibitem{ruiz1985holographic}
F.~Ruiz-Bevia, A.~Celdran-Mallol, C.~Santos-Garcia, J.~Fernandez-Sempere,
  Holographic interferometric study of free diffusion: a new mathematical
  treatment, Applied Optics 24~(10) (1985) 1481--1484.

\bibitem{anand2006diffusivity}
A.~Anand, V.~K. Chhaniwal, C.~Narayanamurthy, Diffusivity studies of
  transparent liquid solutions by use of digital holographic interferometry,
  Applied Optics 45~(5) (2006) 904--909.

\bibitem{he2015development}
M.~He, S.~Zhang, Y.~Zhang, S.~Peng, Development of measuring diffusion
  coefficients by digital holographic interferometry in transparent liquid
  mixtures, Optics express 23~(9) (2015) 10884--10899.

\bibitem{paoletti1997temperature}
D.~Paoletti, G.~S. Spagnolo, Temperature dependence of fluid mixtures
  diffusivity by espi endoscopy, Optics and Lasers in Engineering 26~(4-5)
  (1997) 301--312.

\bibitem{riquelme2007interferometric}
R.~Riquelme, I.~Lira, C.~P{\'e}rez-L{\'o}pez, J.~A. Rayas,
  R.~Rodr{\'\i}guez-Vera, Interferometric measurement of a diffusion
  coefficient: comparison of two methods and uncertainty analysis, Journal of
  Physics D: Applied Physics 40~(9) (2007) 2769.

\bibitem{axelsson}
A. Axelssom, M. Marucci, The use of holographic interferometry and electron speckle pattern interferometry for diffusion measurement in biochemical and pharmaceutical engineering applications, Optics and Lasers in Engineering 46~(12)
  (2008) 865--876.


\bibitem{ambrosini2002speckle}
D.~Ambrosini, D.~Paoletti, A.~Ponticiello, G.~S. Spagnolo, Speckle
  decorrelation study of liquid diffusion, Optics and Lasers in Engineering
  37~(4) (2002) 341--353.

\bibitem{rashidnia2002development}
N.~Rashidnia, R.~Balasubramaniam, Development of an interferometer for
  measurement of the diffusion coefficient of miscible liquids, Applied Optics
  41~(7) (2002) 1337--1342.

\bibitem{torres2012development}
J.~F. Torres, A.~Komiya, E.~Shoji, J.~Okajima, S.~Maruyama, Development of
  phase-shifting interferometry for measurement of isothermal diffusion
  coefficients in binary solutions, Optics and Lasers in Engineering 50~(9)
  (2012) 1287--1296.

\bibitem{spagnolo2004liquid}
G.~S. Spagnolo, D.~Ambrosini, D.~Paoletti, Liquid diffusion coefficients by
  digital moir{\'e}, Optical Engineering 43~(4) (2004) 798--806.

\bibitem{raffel2015background}
M.~Raffel, Background-oriented schlieren (bos) techniques, Experiments in
  Fluids 56~(3) (2015) 60.

\bibitem{settles2017review}
G.~S. Settles, M.~J. Hargather, A review of recent developments in schlieren
  and shadowgraph techniques, Measurement Science and Technology 28~(4) (2017)
  042001.

\bibitem{creath1985phase}
K.~Creath, Phase-shifting speckle interferometry, Applied Optics 24~(18) (1985)
  3053--3058.

\bibitem{takeda1982fourier}
M.~Takeda, H.~Ina, S.~Kobayashi, Fourier-transform method of fringe-pattern
  analysis for computer-based topography and interferometry, JOSA 72~(1) (1982)
  156--160.

\bibitem{kemao2007two}
Q.~Kemao, Two-dimensional windowed fourier transform for fringe pattern
  analysis: principles, applications and implementations, Optics and Lasers in
  Engineering 45~(2) (2007) 304--317.

\bibitem{kemao2004windowed}
Q.~Kemao, Windowed fourier transform for fringe pattern analysis, Applied
  Optics 43~(13) (2004) 2695--2702.

\bibitem{watkins1999determination}
L.~Watkins, S.~Tan, T.~Barnes, Determination of interferometer phase
  distributions by use of wavelets, Optics Letters 24~(13) (1999) 905--907.

\bibitem{watkins2012review}
L.~R. Watkins, Review of fringe pattern phase recovery using the 1-d and 2-d
  continuous wavelet transforms, Optics and Lasers in Engineering 50~(8) (2012)
  1015--1022.

\bibitem{spagnolo1994fourier}
G.~S. Spagnolo, D.~Paoletti, D.~Ambrosini, V.~Bagini, M.~Santarsiero, Fourier
  transform evaluation of digital interferograms for diffusivity measurement,
  Pure and Applied Optics: Journal of the European Optical Society Part A 3~(3)
  (1994) 249.

\bibitem{huntley01}
J. M. Huntley, Automated analysis of speckle interferograms, in P.K. Rastogi (Ed.), Digital speckle pattern interferometry and related techniques, Wiley, 2001.

\bibitem{montresor2016quantitative}
S.~Montresor, P.~Picart, Quantitative appraisal for noise reduction in digital
  holographic phase imaging, Optics Express 24~(13) (2016) 14322--14343.

\bibitem{montresor2019comparative}
S.~Montr{\'e}sor, P.~Memmolo, V.~Bianco, P.~Ferraro, P.~Picart, Comparative
  study of multi-look processing for phase map de-noising in digital fresnel
  holographic interferometry, JOSA A 36~(2) (2019) A59--A66.

\bibitem{xia2018comparative}
H.~Xia, S.~Montresor, P.~Picart, R.~Guo, J.~Li, Comparative analysis for
  combination of unwrapping and de-noising of phase data with high speckle
  decorrelation noise, Optics and Lasers in Engineering 107 (2018) 71--77.

\bibitem{xia2017robust}
H.~Xia, S.~Montresor, R.~Guo, J.~Li, F.~Olchewsky, J.-M. Desse, P.~Picart,
  Robust processing of phase dislocations based on combined unwrapping and
  inpainting approaches, Optics Letters 42~(2) (2017) 322--325.

\bibitem{gao2009real}
W.~Gao, N.~T.~T. Huyen, H.~S. Loi, Q.~Kemao, Real-time 2d parallel windowed
  fourier transform for fringe pattern analysis using graphics processing unit,
  Optics Express 17~(25) (2009) 23147--23152.

\bibitem{van2016real}
S.~Van~der Jeught, J.~J. Dirckx, Real-time structured light profilometry: a
  review, Optics and Lasers in Engineering 87 (2016) 18--31.

\bibitem{wang2019fast}
H.~Wang, H.~Zeng, P.~Chen, R.~Liang, L.~Jiang, Fast single fringe-pattern
  processing with graphics processing unit, Applied Optics 58~(25) (2019)
  6854--6864.

\bibitem{vishnoi2019rapid}
A.~Vishnoi, G.~Rajshekhar, Rapid deformation analysis in digital holographic
  interferometry using graphics processing unit accelerated wigner--ville
  distribution, Applied Optics 58~(16) (2019) 4420--4424.

\bibitem{hayes2009statistical}
M.~H. Hayes, Statistical digital signal processing and modeling, John Wiley \&
  Sons, 2009.

\bibitem{stoica2005spectral}
P.~Stoica, R.~L. Moses, Spectral analysis of signals, Pearson Prentice Hall
  Upper Saddle River, NJ, 2005.

\bibitem{golub2012matrix}
G.~H. Golub, C.~F. Van~Loan, Matrix computations, Vol.~3, JHU press, 2012.

\bibitem{chapra2012applied}
S.~C. Chapra, Applied numerical methods, McGraw-Hill Columbus, 2012.

\bibitem{herraez2002fast}
M.~A. Herr{\'a}ez, D.~R. Burton, M.~J. Lalor, M.~A. Gdeisat, Fast
  two-dimensional phase-unwrapping algorithm based on sorting by reliability
  following a noncontinuous path, Applied Optics 41~(35) (2002) 7437--7444.

\bibitem{nvidia2011nvidia}
C.~Nvidia, Nvidia cuda c programming guide, Nvidia Corporation 120~(18) (2011)
  8.

\bibitem{sanders2010cuda}
J.~Sanders, E.~Kandrot, CUDA by example: an introduction to general-purpose GPU
  programming, Addison-Wesley Professional, 2010.

\bibitem{gao2012parallel}
W.~Gao, Q.~Kemao, Parallel computing in experimental mechanics and optical
  measurement: a review, Optics and Lasers in Engineering 50~(4) (2012)
  608--617.

\bibitem{wang2018parallel}
T.~Wang, Q.~Kemao, Parallel computing in experimental mechanics and optical
  measurement: A review (ii), Optics and Lasers in Engineering 104 (2018)
  181--191.

\bibitem{van2011numpy}
S.~Van Der~Walt, S.~C. Colbert, G.~Varoquaux, The numpy array: a structure for
  efficient numerical computation, Computing in Science \& Engineering 13~(2)
  (2011) 22.
  
\bibitem{crank}
J. Crank, The Mathematics of Diffusion, 2nd ed., Oxford University Press, 1975.
  

\end{thebibliography}

\end{document}